\documentclass[10pt,twocolumn]{article}

\usepackage[utf8]{inputenc}
\usepackage[T1]{fontenc}
\usepackage[margin=0.75in]{geometry}
\usepackage{graphicx}
\usepackage{amsmath,amssymb}
\usepackage{booktabs}
\usepackage{array}
\usepackage{makecell}
\usepackage{url}

\providecommand{\Description}[1]{}

\title{Verification and Validation (V\&V)-in-the-Loop for RISC-V Design: The Holistic Vision of BZL}

\author{%
Sajjad Ahmed \and
Alexander Kropotov \and
Roberto Ignacio Genovese \and
Bernat Homs \and
Eloi Merino \and
Francesco Urbani \and
Henrique Yano \and
Iván Díaz \and
Joan Gracia Fernandez \and
Matteo Toselli \and
Muhammad Imran \and
Muhammad Abu Bakar Umar Haider Iqbal \and
Nadeem Yaseen \and
Quswar Abid \and
Shaista Cheema \and
Samuel Sanchez \and
Daniel Garcia \and
Joan Cabré \and
Mostafa Elyasi \and
Fernando Ayats \and
Miquel Moreto \and
Teresa Cervero \and
Oscar Palomar \and
Behzad Salami\\[0.5em]
\normalsize Barcelona Supercomputing Center (BSC)
}
\date{}

\begin{document}

\maketitle

\begin{abstract}
The Barcelona Zetascale Lab (BZL) project aims to strengthening Europe's capacity in the design and manufacture of RISC-V based high-performance computing chips. In this context, we present a holistic pre-silicon verification and validation (V\&V) methodology targeting highly robust RISC-V chip designs. This paper provides an overview of BZL's V\&V approach, which integrates three complementary platforms: (1) a UVM-based verification environment to thoroughly validate RTL functionality; (2) an FPGA-based validation platform that enables system-level pre-silicon hardware-software RTL validation; and (3) a CI/CD flow that continuously automates build, deployment, and tests across these domains. By embedding these platforms into an industrial-grade V\&V loop and exploiting large-scale CPU and FPGA hardware infrastructures, the BZL project enables continuous evolution of reliable hardware development and software integration. We believe that the BZL's V\&V flow represents a robust and scalable foundation for ensuring the pre-silicon functional correctness and system-level validation of RISC-V chip designs, and can serve as a key enabler for strategic initiatives in Europe, such as EPI and DARE, and beyond.
\end{abstract}


 \begin{figure}[hb]
  \centering
  \includegraphics[width=\linewidth]{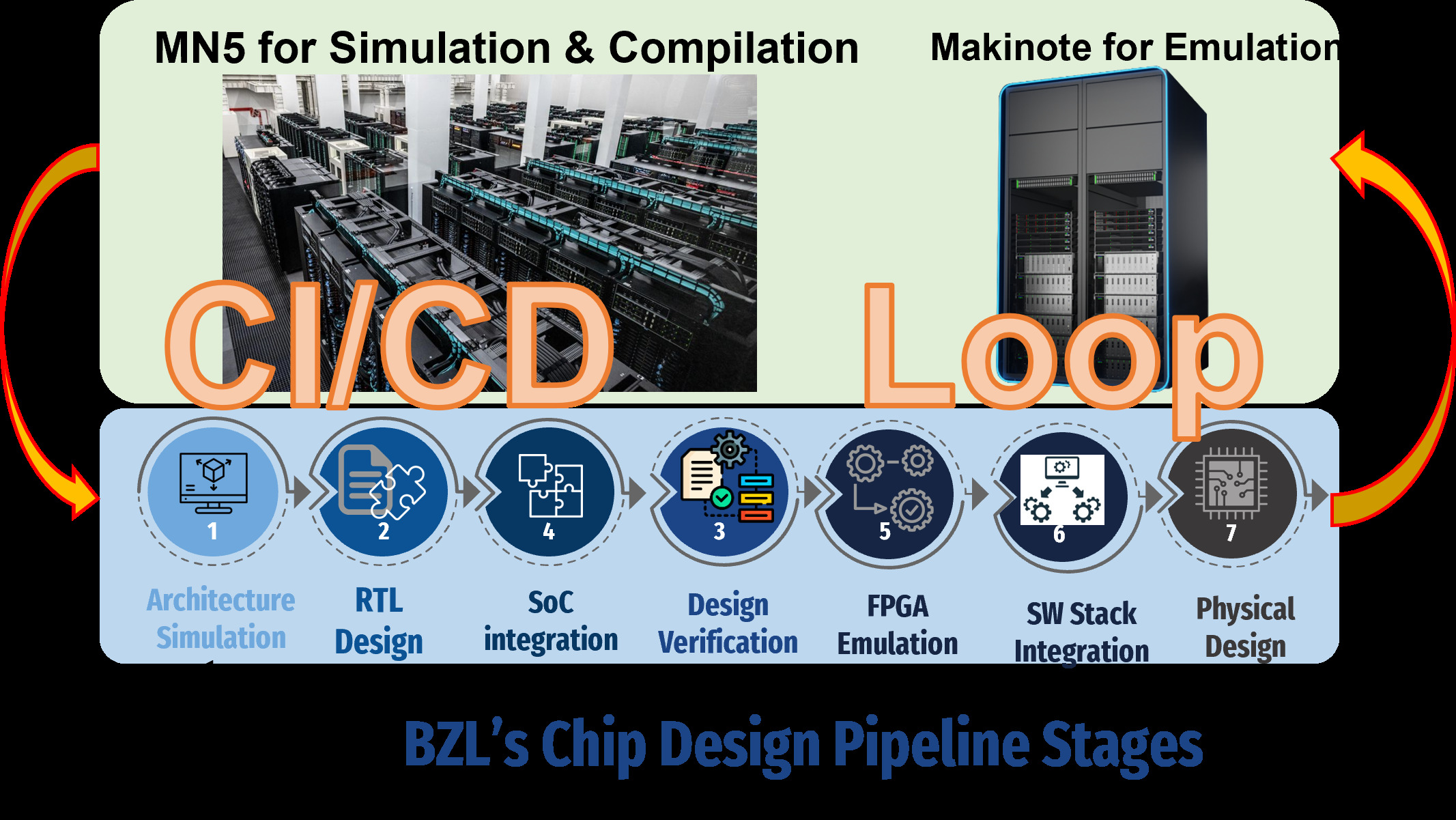}
  \caption{Overview of BZL's Chip Design Flow: Due to time and space limit, this paper will focus on some selected stages, i.e., Design Verification and System Validation.}
  \label{fig:vv}
\end{figure}

\section{Introduction}
 %ADD YOUR INPUT HERE   
 \label{sec:intro}

The \textbf{Barcelona Zetascale Lab (BZL)}~\footnote{https://bzl.es/en} is a three-years, large-scale, and collaborative research project led by the Barcelona Supercomputing Center (BSC), started on June 2023. The project aims to strengthen Europe’s technological sovereignty in \textbf{RISC-V–based HPC chips} by establishing an open and end-to-end design ecosystem spanning architecture simulation, RTL development, pre-silicon verification and validation (V\&V), Physical Design, and software stack integration. One of the key challenges addressed by BZL is the growing cost and complexity of pre-silicon V\&V, which is exacerbated by the extensibility of the RISC-V ISA and the increasing scale of modern SoCs. To address this issue, BZL proposes a \textbf{holistic “V\&V-in-the-Loop” methodology} that tightly integrates functional verification, system-level validation, and CI/CD automation into a continuous and scalable workflow and by effectively exploiting the large-scale CPU and FPGA infrastructures in the center (see Fig.~\ref{fig:vv}).

Pre-silicon Verification and validation (V\&V) is one of the most resource-intensive phases in the chip development flow, often consuming over 60–70\% of the overall effort~\cite{ver1}. Aiming robustness in the complex multi-core RISC-V based RTL design flow, the BZL project adopts a continuous and scalable V\&V strategy. This infrastructure is built around three tightly coupled components:

\begin{itemize}
\item \textbf{UVM-Based RTL Verification:} A reusable and hierarchical UVM environment to validate the multi-core RISC-V SoC with vector acceleration (RVV 1.0)~\cite{vpudvpaper}. %Verification scales from unit-level vector accelerator~\cite{vpudvpaper} testing to full CPU subsystem validation~\cite{dvposter}, using co-simulation with the Spike ISS as a golden reference and coverage-driven regressions.
\item \textbf{FPGA-Based System-Level Validation:} Towards a complementary system-level validation of the underlying SOC, we map in onto FPGAs using large-scale cluster \textit{Makinote}~\cite{perdomo2024makinote}. Complementary tests in this stage includes a large-set of bare-metal workloads, Linux boot, driver validation, and realistic software stacks. %The platform supports multi-FPGA partitioning, early OS bring-up, and hardware/software co-validation at speeds unattainable with simulation alone.
\item \textbf{CI/CD Automation:} A unified CI/CD infrastructure orchestrates the whole V\&V flow, e.g., linting, RTL simulation, UVM regressions, FPGA bitstream generation, deployment, and large-scale test execution. %Validation workloads are distributed across multiple FPGAs to significantly reduce turnaround time and enable continuous hardware validation.
\end{itemize}

The RTL design developed within the BZL project is an in-house NoC-based, multi-core out-of-order (OoO) scalar processor capable of booting Linux and supporting RVV 1.0 through a BSC-designed Vector Processing Unit (VPU) \cite{minervini2023vitruvius}. Due to space limit, in this paper, we don not cover the description of this architecture, neither the fabrication and silicon prototype characteristics.

The BZL project demonstrates that a scalable V\&V framework is essential for modern complex SOC design. Many integration and software–hardware interaction issues cannot be uncovered through simulation alone. Beyond its technical results, BZL establishes a reusable and collaborative blueprint for future European chip initiatives. In this paper, we will introduce our methodology and provide a practical overview of our framework and lessons learned, highlighting key innovations and representative results, as outlined below:

\begin{itemize}
\item A scalable, CI-driven FPGA validation platform equipped with advanced features: a fully cycle-accurate SoC emulation and FPGA-level peripheral integration, integrated extensive bare-metal and Linux-based test suites, effectively exploiting large number of FPGAs to parallelize the tests and support large SOCs partitioned onto multiple FPGA, and fully automated RTL-to-FPGA regression with bitstream generation, execution, and analysis.
%\item Automated execution of more than \textbf{1,700 system-level tests}, including ISA compliance, vector benchmarks, memory-consistency litmus tests, and OS-level stress tests.
%\item Up to \textbf{8$\times$ reduction in system-level validation time} through parallel FPGA execution.
%\item High functional confidence supported by stable regression results and consistent coverage metrics across RTL revisions
\item A scalable, CI-driven, and industrial-grade verification framework combining a hierarchical and reusable UVM stack (from vector unit to multi-core subsystem) equipped with advance features: advanced constrained-random verification mechanism, Spike ISS co-simulation with retirement-level architectural scoreboarding, advanced constrained-random corner-case testing, multi-core subsystem integration (NoC, caches, AXI, DMA, etc.).
%\item Development of a hierarchical, reusable UVM verification stack (vpu-dv → core-uvm → cpu-subsystem-uvm) enabling scalable validation from unit-level vector accelerator to full dual-core subsystem, providing a \textbf{true systematic vertical reuse across abstraction levels}. This avoids testbench duplication, re-verification effort explosion, interface re-modeling errors.
%\item Integration of Spike ISS co-simulation as a golden reference with step-and-compare ISA scoreboarding at retirement level, ensuring architectural equivalence across scalar and vector pipelines and providing a \textbf{cycle-aware architectural equivalence framework}.
%\item Advanced constrained-random verification, including random pipeline kills (branch misprediction modeling), randomized page faults and vector memory exceptions, interrupt synchronization between DUT and reference model via DPI, providing \textbf{corner-case micro-architectural robustness} which goes beyond compliance testing.
%\item Multi-core UVM verification with NoC, L2 cache, AXI fabric, DMA, PLIC, CLINT, and JTAG VIP integration for full subsystem validation.
%\item Deployment of an industrial-grade, CI-integrated UVM regression framework enabling automated merge-request validation, coverage-driven weekly regressions, large-scale test orchestration, and continuous RTL-to-FPGA verification, ensuring early bug detection, regression stability, and reproducible hardware validation across design revisions.
\end{itemize}

\section{UVM-Based RTL Verification}   
 %ADD YOUR INPUT HERE   
\label{sec:uvm}
Our functional verification methodology is rooted in the Universal Verification Methodology (UVM) and applies a hierarchical "bottom-up" approach. We have developed three distinct, reusable, and progressively complex UVM environments that mirror the BZL design's integration flow: (1) \texttt{vpu-dv} for the Vector Processing Unit (VPU), (2) \texttt{core-uvm} for the scalar core, and (3) \texttt{cpu-subsystem-uvm} for the integrated dual-core subsystem \cite{dvposter}. All three environments leverage co-simulation with Spike, the RISC-V Instruction Set Simulator (ISS), as the golden reference model\footnote{https://github.com/riscv-software-src/riscv-isa-sim}.

\subsection{VPU-DV: Vector Accelerator Verification}
The foundational environment, \texttt{vpu-dv}, targets the VPU, a vector co-processor implementing the RISC-V V-extension (RVV-1.0). Verification at this level focuses on two distinct targets: the correctness of the RVV implementation and the fidelity of the custom Barcelona Vector Interface (BVI) protocol, which connects the VPU to a scalar core.

The environment is built with a common, reusable \texttt{vpu-dv} repository and a project-specific layer (\texttt{epac2-vpu-dv}) that instantiates the BVI components \cite{vpudvpaper}. In its "Active" mode, the testbench's UVM Verification Components (UVCs) emulate the scalar core. A BVI agent, driven by instructions processed by Spike, manages all BVI sub-interfaces to generate legal vector stimuli.

This stimulus generation is comprehensive, covering not only standard arithmetic and memory operations but also advanced protocol-level scenarios. The agent injects random "kills" to mimic branch mispredictions, forcing the VPU to discard instructions in its pipeline. It also models random exceptions, such as page faults during vector memory operations, testing the VPU's ability to halt, report the exception, and correctly resume execution.

Checkers include an ISA scoreboard that compares the VPU's architectural state against Spike's, and a protocol scoreboard that monitors BVI-specific transactions to ensure adherence to the interface specification.

\subsection{Core-UVM: Scalar Core Verification}
The next level of integration is \texttt{core-uvm}, which verifies the BZL's custom complete 64-bit scalar core with its integrated VPU. This environment is a significant step up in complexity as it validates the entire instruction pipeline, memory hierarchy, and interrupt handling of the core itself.

While \texttt{vpu-dv} actively emulates the scalar core, \texttt{core-uvm} instantiates the full RTL core and its associated \texttt{vpu-dv} environment in "Passive" mode. The core now executes compiled RISC-V binaries (generated by tools like \texttt{riscv-dv}\footnote{https://github.com/chipsalliance/riscv-dv}), and the testbench is responsible for modeling the rest of the system.

Key components of this environment include:
\begin{itemize}
    \item \textbf{Cache Agents:} The \texttt{icache\_agent} and \texttt{dcache\_agent} act as reference models for the instruction and data memory hierarchies, respectively. They service cache requests from the core's interfaces, introducing randomized delays to stress the pipeline.
    \item \textbf{Interrupt Agent:} This agent randomly generates and drives external, timer, and software interrupts to the core's interrupt interface. It simultaneously notifies the Spike reference model of the same interrupts via DPI calls, ensuring both DUT and reference model operate on a synchronized view of external events.
    \item \textbf{Step-and-Compare Logic:} An instruction manager processes instructions as they are completed (retired) by the DUT. An ISA scoreboard then performs a detailed "step-and-compare" check, validating the architectural state (PC, registers, CSRs, trap status) of the retired instruction against the golden state from Spike.
\end{itemize}

\subsection{CPU-Subsystem-UVM: Multi-Core Integration Verification}
The final and most complex environment is \texttt{cpu-subsystem-uvm}, which targets the full dual-core CPU subsystem. This testbench integrates two CPU cores, their respective \texttt{core-uvm} instances, a Network-on-Chip (NoC), L2 cache, and various system-level peripherals.

The primary challenge at this level shifts from single-core correctness to verifying multi-core interaction and system-level functionality. The environment builds upon \texttt{core-uvm} by instantiating two complete core verification stacks and adding UVCs for all subsystem-level components, including:
\begin{itemize}
    \item \textbf{AXI Agents:} An AXI-Slave memory model emulates the main memory and bootrom, while an AXI-Master VIP is used to verify the system's DMA bridge.
    \item \textbf{Peripheral VIPs:} Dedicated agents and VIPs drive and monitor the JTAG/debug interface and the subsystem-level PLIC and CLINT interrupt controllers.
\end{itemize}

\subsection{Regression Testing and CI Infrastructure}
The three UVM environments are integrated into a comprehensive CI infrastructure built on Gitlab pipelines. This CI system automates our entire verification flow, with regressions automatically triggered for Merge Requests (MRs), as well as on daily and weekly schedules. The weekly pipelines are primarily responsible for gathering and reporting detailed code and functional coverage.

This continuous regression suite executes a wide array of tests to ensure thorough validation at all levels, running tests on a per-core and a multi-core basis. The test suites include:
\begin{itemize}
    \item \textbf{Random Tests:} Heavily constrained random tests generated with \texttt{riscv-dv}, covering scalar-only, memory-intensive, bare-metal (physical), and virtual-memory configurations.
    \item \textbf{ISA Tests:} Directed tests targeting specific scalar (privileged and unprivileged) and vector instruction functionality.
    \item \textbf{Directed Peripheral Tests:} Specific test cases for verifying the DMA (via AXI), the PLIC and CLINT interrupt controllers, and the debug module (via JTAG).
    \item \textbf{Benchmarks:} A wide variety of benchmarks, including \texttt{matmul}, \texttt{median}, and \texttt{axpy}, as well as the 5Js suite (e.g., \texttt{gemm}, \texttt{spmv}, \texttt{stream}) with varying data sizes.
    \item \textbf{Compliance and Corner-Case Tests:} The \texttt{riscof} compliance suite, Litmus tests for memory consistency, and X-propagation tests to ensure design robustness.
\end{itemize}

This rigorous, automated process yields high confidence in the design's correctness. At the top level of integration, our weekly regressions consistently achieve high code coverage, with typical results of \textbf{91\% Statements}, \textbf{82\% Branches}, and \textbf{65\% Toggle}, for a \textbf{Total Coverage of 80\%} (without waivers or exclusions).

% --------------------------------------
\begin{figure}
  \centering
  \includegraphics[width=0.8\linewidth]{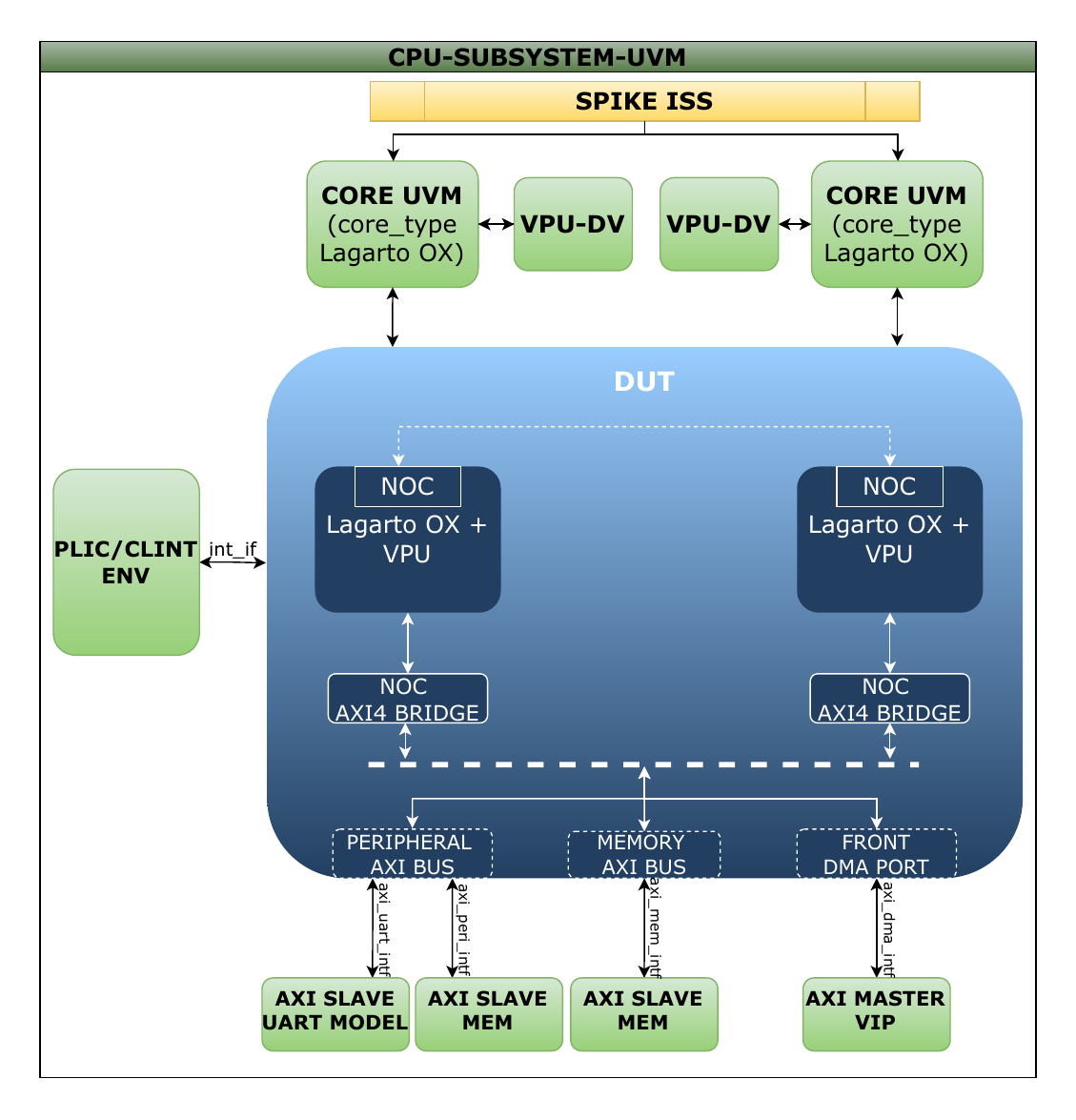}
  \Description{Diagram of the CPU-Subsystem-UVM verification environment showing the components and connections used for FPGA testing and validation.}
  \caption{CPU-Subsystem-UVM verification environment.}
  \label{fig:cpu-ss-uvm}
\end{figure}
% ---------------------------------

\section{FPGA-Based System Validation}     %ADD YOUR INPUT HERE   
\label{sec:fpga}

\begin{figure*}
  \centering
  \includegraphics[width=0.8\textwidth]{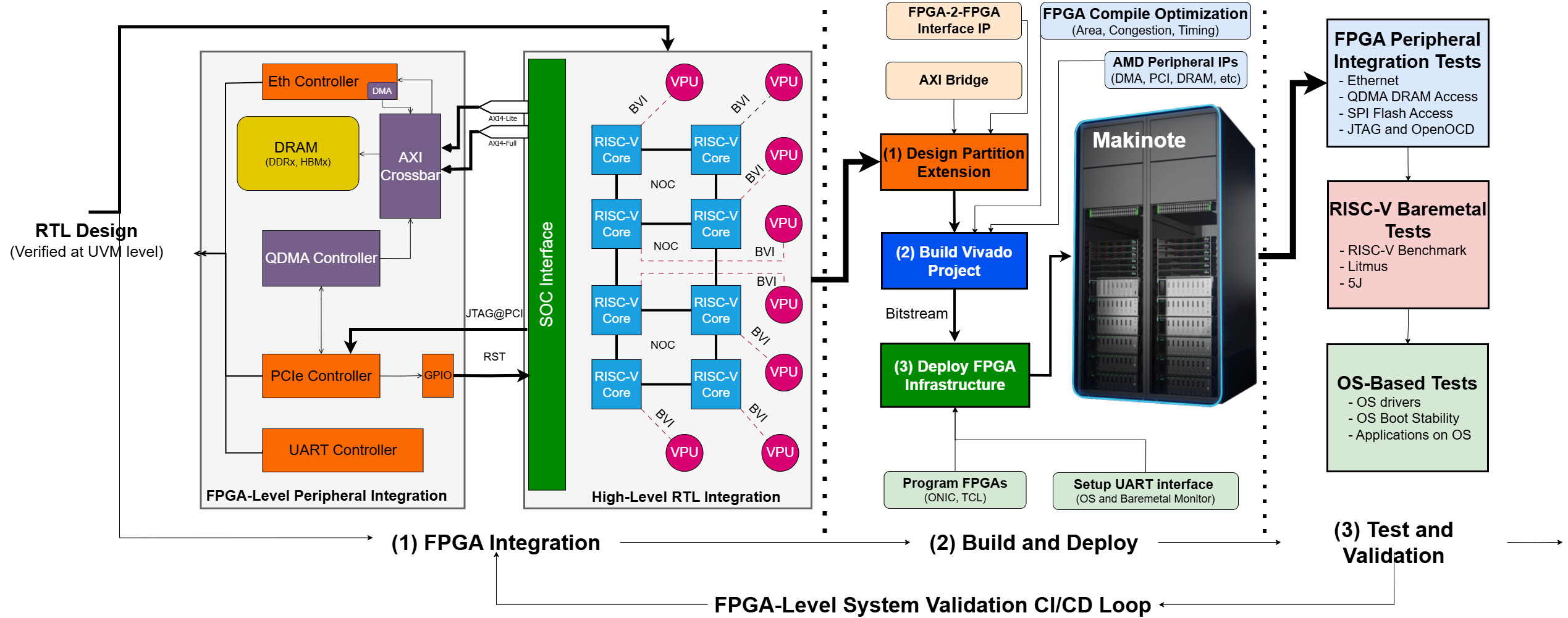}
  \caption{FPGA-Level System Integration, Build and Deploy, and Test and Validation}
  \Description{Diagram illustrating the FPGA-level workflow, including system integration, build and deployment process, and subsequent test and validation stages.}
  \label{fig:FPGA}
\end{figure*}

After RTL design passes the UVM simulation tests, we perform the complementary system validation tests at FPGA level. This section outlines the FPGA infrastructure used for the pre-silicon system validation of BZL´s RISC-V based design. Our framework is structured into three primary phases: \textit{FPGA Integration}, \textit{Build and Deploy}, and \textit{Test and Validation}, as shown in Fig~\ref{fig:FPGA}. This methodology effectively bridges the gap between RTL simulation and silicon bring-up, enabling complementary system validation, early integration of OS and software stack, performance profiling, and architectural exploration.

\subsection{FPGA Integration}
\label{subsection:fpga1}

In this phase, the RTL design is connected with FPGA-level infrastructure and platform peripherals to form a complete software-bootable system. This includes interfacing DDR/HBM memory controllers, Ethernet for external accesses, UART for console access, PCIe DMA interfaces for host communication, SPI to support non-volatile memory for firmware booting and debug interfaces such as JTAG. Our FPGA emulation platform is built upon AMD/Xilinx peripheral IPs. Standard AXI interfaces are used across the system, allowing seamless connection between the compute complex and FPGA peripheral subsystem while preserving the original SoC architecture. Integration and configuration steps are handled through automated scripts and flows that stitch the RTL with platform IPs and prepare the FPGA environment for execution. This automation minimizes manual effort, ensures repeatability, and allows rapid iterations while maintaining consistency with the intended SoC behavior. The earlier version of the toolset is available in public\footnote{https://github.com/bsc-loca/fpga-shell}.

\begin{table*}[!ht]
\centering
\caption{Full Test Execution Summary on Single vs. 8 FPGAs}
\footnotesize % helps fit in dual columns
\setlength{\tabcolsep}{2.5pt} % reduce column spacing
\begin{tabular}{lclccc}
\hline
\textbf{Tests Category} & \textbf{\shortstack{Name}} & \textbf{\shortstack{Description}} & \textbf{\shortstack{Total Tests}} & \textbf{\shortstack{Execution Time \\(1 FPGA)}} & \textbf{\shortstack{Execution Time \\ (8 FPGAs)}} \\
\hline

\textbf{\shortstack{FPGA Integration Tests}} & \texttt{spi} &Flash tests (Read/Write/Erase) & 1 & 59s & -- \\
                      & \texttt{jtag-debug}      & OpenOCD and JTAG tests (mem/reg) & 109 & 1102s & $\sim$138s \\

\hline
 & \texttt{vec-axpy}      & Vector fused multiply-add & $\sim$50  & 524s   & $\sim$65.5s \\
                          & \texttt{vec-gemm}      & Dense matrix multiplies, multiple sizes & $\sim$70  & 728s &   $\sim$91s \\
                          & \texttt{vec-stream}    & bandwidth, load/store, copy, triad & $\sim$50  & 536s  &  $\sim$67s \\
 \textbf{\shortstack{RISC-V Baremetal Tests}}                         & \texttt{vec-somier}    & Reduction and exponential kernels & $\sim$20  & 208s   & $\sim$26s \\
                          & \texttt{spmv}          & Sparse matrix-vector multiply & $\sim$40  & 415s   & $\sim$53.5s \\
                          & \texttt{litmus}        & Multi-core coherency tests  & 1370      & 13892s & $\sim$1736.5s \\

                            & \texttt{rv-tests}     & RISC-V Benchmark  & 18   & 132s  & $\sim$16.5s \\

\hline
        & \texttt{ethernet-driver}  & (ping/scp/ssh) fpga-to-fpga, host-to-fpga& 6   & 541s & $\sim$91.6s \\
                           & \texttt{linux-boot} (8 runs)      & Buildroot with OpenSBI & 1  & 115s & -- \\
      \textbf{OS-Level Tests}                       & \texttt{stress-ng} (8 runs)       & system stress tests  & $\sim$500  & 670s & -- \\
                           & \texttt{test\_dd} (8 runs)        & Linux dd command & 1  & 22s & -- \\
                           & \texttt{test\_plic} (8 runs)      & interrupt tests & 1  & 18s & -- \\

\hline
\textbf{Total} & \texttt{all-tests} & --- & 1738 & 18754s ($\sim$5.2h) & $\sim$3169.6s ($\sim$53m) \\
\hline
\end{tabular}
\label{tab:fpga_parallel_results}
\end{table*}

\subsection{Build and Deploy}
\label{subsection:fpga2}

After the SoC RTL is integrated with the FPGA platform, the next step is to prepare the design for FPGA execution, bitstream generation and bring up the hardware environment for bare-metal and OS testing. This phase adds required project configurations including design partition extension, Vivado project creation and deploy FPGA infrastructure through automated flow based on TCL/Python scripts. The hardware FPGA platform is BSC’s large-scale FPGA cluster, Makinote~\cite{perdomo2024makinote}. Makinote is composed of 96 Alveo u55c FPGA boards formed into 12 FPGA nodes, each hosting 8 FPGAs, connected through PCIe to the node. Also, all FPGAs have direct FPGA-2-FPGA connection and are interconnected through Ethernet switches as well. We exploit this platform and take benefit of large number of FPGAs in two ways: first, to support the emulation of large SoC designs through RTL partitioning across multiple-FPGAs, elaborated below. Second, to distribute tests across several FPGAs to reduce the test time, elaborated in~\ref{subsection:fpga3}.

We elaborate each step of this phase as below:
\begin{itemize}
    \item \textbf{Design Partition Extension}: For designs that exceed the capacity of a single FPGA, the framework supports seamless scaling to a multi-FPGA environment. In this model, the SoC is partitioned across multiple devices, with additional components such as AXI bridges, serializer-deserializer links, and synchronization logic added to maintain functional connectivity between partitions. While this approach enables emulation of very large SoCs, it introduces trade-offs including increased communication latency across FPGA boundaries and higher integration complexity. Nevertheless, it provides a scalable path for validating large-core systems that cannot fit in a single FPGA. 
\item \textbf{Build Vivado Project}: Once the design topology is finalized, an automated TCL-based flow builds the Vivado project. Key actions in this phase include:
Import SoC RTL and FPGA peripheral IP (DDR, PCIe DMA, UART, Ethernet, timers, etc)
Generate the block design wrapper
Define memory maps and system address space
Apply board-level files (device config, clocks, reset, IOs)
Apply design constraints
Configure synthesis and implementation runs (with optimization settings for Area, Congestion, and Timing)
Launch synthesis and bitstream generation flow.
\item \textbf{Deploy FPGA Infrastructure}: Finally, after the bitstream(s) generated through the flow, we deploy the hardware in the FPGA cluster, through programing the FPGAs and setting up the necessary interfaces for users. Communication with the host system is established via PCIe for uploading test binaries and operating system images, UART for interacting with the OS console, and OpenOCD for debugging. All processes are automated through scripts, which accelerates the development cycle, making it straightforward to rebuild and test updated RTL versions, experiment with new designs, and quickly address any bugs.
\end{itemize}

\subsection{Test and Validation}
\label{subsection:fpga3}
Our validation strategy progresses from basic hardware checks to full system-level execution. Once the FPGA is programmed, initial hardware diagnostics confirm the basic functionality of the FPGA system, e.g., memory, high-speed communication links, and debug interfaces. Subsequently, we execute a comprehensive test suite for system validation, categorized in three classes: 
\begin{enumerate}
    \item FPGA-Level Integration Tests: to verify the FPGA-RTL integration.
    \item RISC-V Baremeral Tests: A set of tests to verify the functionality of RTL with RISC-V based benchmark suites at Baremetal (without OS), e.g., axpy\footnote{https://github.com/RALC88/riscv-vectorized-benchmark-suite}, gemm\footnote{https://github.com/danieldk/gemm-benchmark}, stream\footnote{https://github.com/jeffhammond/STREAM}, somier\footnote{https://github.com/moimfeld/rivec}, spmv\footnote{https://bebop.cs.berkeley.edu/spmvbench/}, litmus\footnote{https://github.com/LITMUS-Benchmark-Suite/}.
    \item OS-Based Tests: to validate the Driver, OS, and software stack, e.g., stress-ng benchmark suite\footnote{https://github.com/ColinIanKing/stress-ng}.
\end{enumerate}

To accelerate system validation, we developed an automated flow that executes the test suite in parallel across 8 FPGAs within a single Makinote, in which hosts 8 Alveo U55c FPGAs. When tests are executed sequentially on one FPGA, the complete suite of 1,738 tests requires approximately 18,754 seconds (~5.2 hours) to finish, as shown in table. \ref{tab:fpga_parallel_results}. This table contains the execution time of the BZL's RTL design, a dual-core with VPU equipped running with 20mhz on FPGAs. 

By distributing test execution across 8 FPGAs in a node, we reduce the total execution time to ~3,170 seconds (~53 minutes). This represents a ~6× overall speed-up. Certain test categories show near-linear speed-up, since they are independent and can run on each FPGA. For example, the RISC-V benchmark suite reduces from 132 s to ~16.5 s, and vector workloads such as VEC-GEMM drop from 728 s to ~91 s. Larger sets of micro-architecture Litmus tests also observe significant improvement, decreasing from 13,892 s to ~1,736 s. However, there are tests that are unified and cannot be divided and distributed, e.g., Linux boot. For such scenarios, we repeat the execution 8 times, for ensuring the stability of the test.

\section{CI/CD Automation}
 %ADD YOUR INPUT HERE  
 \label{sec:ci}
Automation through continuous integration and continuous delivery (CI/CD) is a well-known and common practice in software engineering, but it is still not very mature in hardware design. In hardware design, CI/CD refers to the automated process of integrating design changes, running simulations and checks, and delivering verified hardware descriptions or FPGA bitstreams, enabling faster iterations, early error detection, and consistent design quality throughout the development cycle. With the leverage of CI/CD for open-source software environments, there is a wide range of frameworks and solutions available, but the integration of verification tools and components are predominantly custom solutions built around GitHub Actions, GitLab CI, or Jenkins, which is widespread for test servers in industry. The advantage of these solutions is that they are built on well-established infrastructure software
 and adopt modern software engineering methods of CI/CD, building on existing tool collections and verification flows, and with well-established automation frameworks such as Gitlab Runners. These enhancements increase engineering productivity while lowering the learning curve, increasing process reproducibility and traceability, infrastructure scalability, and confidence in the stability of the design.

We aims to apply similar approach for hardware design. Enabling the CI infrastructure brings the possibility of increasing productivity for all hardware design and development teams, since there is no need for them to master the complexity of the whole flow. The experts are focused on their area of expertise, by reading and analyzing the results from the CI flow, and applying the required changes in the hardware design code. 

In BZL prject, we created continuous integration (CI) pipelines for linting, verification, simulation, and design emulation. This comprises infrastructure for gathering functional correctness, reporting coverage, and performance metrics, automated regression environments, and automated deployments for running long and complex testbenches (e.g. booting an operating system). In addition, current flows incorporate functional verification techniques and transform them into repeatable CI/CD pipelines, spanning abstraction levels from coding to netlists. The physical simulation is out of the scope of this paper.

\subsection{BZL's CI setup}
The CI pipeline validates hardware RTL and system-level functionality across multiple stages within the digital design flow, from static analysis to full FPGA emulation. Automating the flow ensures design integrity, regression control, and continuous hardware validation. Therefore, the main objectives of enabling the CI are the following: 
\begin{itemize}
    \item \textbf{Early detection} of syntax, lint, or structural issues.
    \item \textbf{Functional verification} through simulation and UVM regressions.
    \item \textbf{Hardware validation} using FPGA emulation with parallel test execution 
\end{itemize}

% --------------------------------------
\begin{figure}
  \centering
  \includegraphics[width=0.8\columnwidth]{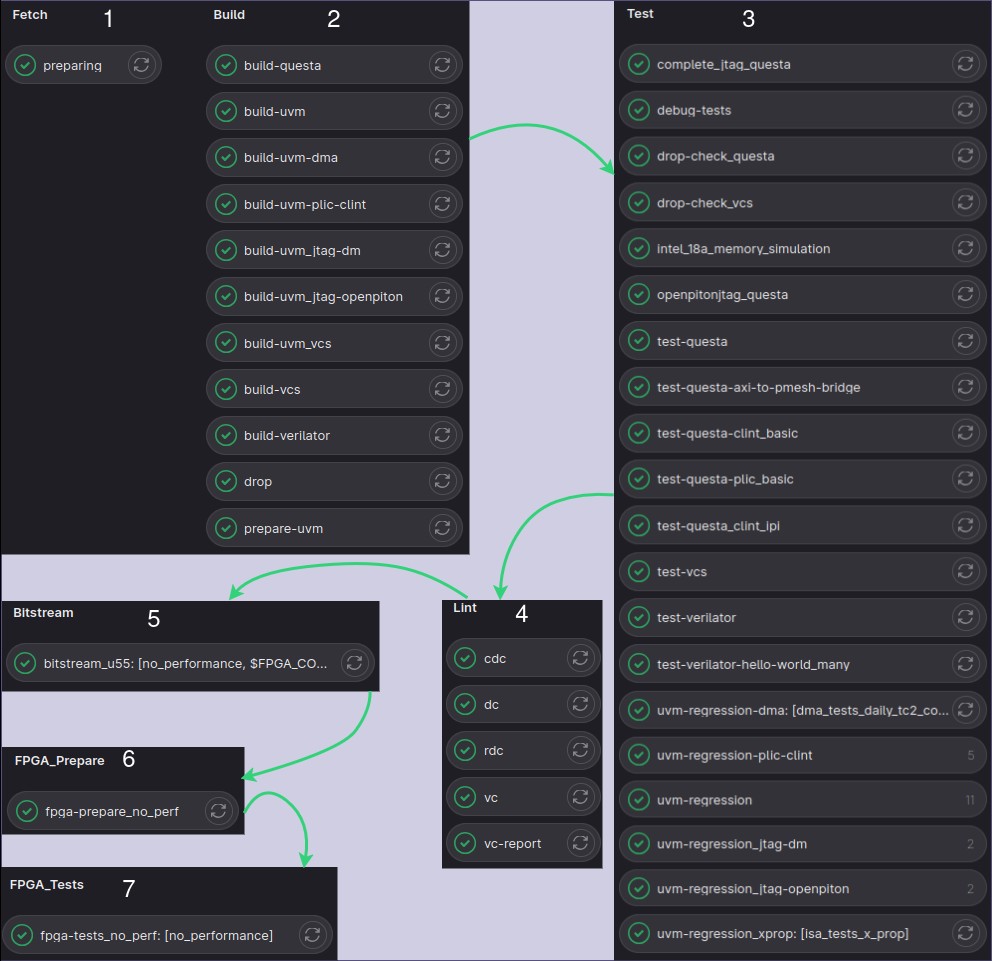}
  \Description{Diagram of CI pipelines used to automate the verification and validation unified flow for FPGA tests.}
  \caption{CI pipelines for automating the verification and validation unified flow.}
  \label{fig:pipeline}
\end{figure}
% ---------------------------------
To achieve these goals, our CI infrastructure groups the tests (described in Sections \ref{sec:uvm} and \ref{sec:fpga}) in the following categories, as shown in Fig. \ref{fig:pipeline}: 
\begin{itemize}
    \item \textbf{Linting and Structural Checks}: Static analysis (VC static, CDC, RDC, DC elaboration).
    \item \textbf{Hardware Simulation Tests}: Functional tests in Questa, Verilator, and VCS, covering ISA, CLINT/PLIC, and custom directed tests.
    \item \textbf{UVM Verification}: Full regression environment with variants (DMA, JTAG, xprop)
    \item \textbf{FPGA Emulation}: Hardware-level validation using FPGA clusters with up to 8 boards. Depending on the purpose of the emulation and to maximize the hardware utilization, different jobs have been configured, which triggers by: 
    \begin{itemize}
        \item \textbf{Merge request runs}: Run a standard functional suite.
        \item \textbf{Daily runs}: Regression tests, a subset of performance tests, and RVV tests.
        \item \textbf{Weekly runs}: there are two set of runs under this category: 1. executing a full performance suite, and 2. running performance validation tests.
    \end{itemize}
    \item \textbf{Drops and Packaging}: Reproducible bundles and post-build checks (drop-checks).
\end{itemize}

\subsection{CI Schedulers and Execution Rules}
Different schedulers determine which of the previous job subset runs. To simplify the maintenance and configuration of these schedulers, job's triggering are controlled via variables, labels, and commits' messages. Therefore, 5 different configurations might be activated:
\begin{itemize}
    \item \textbf{Torture or Per-Commit}: Quick feedback for main branch MRs or commits tagged with \textit{verification}. Runs lint, smoke simulations, and selective UVM builds.
    \item \textbf{Daily}: Activated by variable and run standard linting, full simulation, bitstream generation, and daily FPGA tests.
    \item \textbf{Weekly}: Triggered with labels. This configuration includes performance configurations and stability tests (8 FPGA cluster).
    \item \textbf{Stability}: Uses the variable \textit{stability\_test} to run extended\-duration tests pinned to a specific commit SHA.
    \item \textbf{Disable Controls}: Some specific labels (e.g., \textit{ci-test}, \textit{disable-uvm}, or \textit{no-bitstream-gen, among others} allow selective exclusions.
\end{itemize}

\subsection{CI Impact in FPGA Emulation}
The FPGA-based design emulation is a critical stage within the chip design flow for validating the design. In our project, including this stage as part of the CI flow has helped system-level validation. Moreover, the CI has helped to find, identify, and solve stability issues in the design. 
The FPGA-based emulation uses Alveo U55C FPGA boards for deployment. In the case of the testing the stability of the design, we exploit multiple FPGAs in parallel. 

These jobs are configured to automatically load bitstreams, deploy them on the target FPGA, run tests in baremetal, and/or boot Linux, i.e., automating the FPGA flow described in Section \ref{sec:fpga}. All these executions are monitored, and results are aggregated to YAML reports. In order to contemplate failure scenarios, each test has been configured with a corresponding \textit{Failure threshold}, which is configurable. This process is triggered with every merge request to the main branch, as part of the integration repository where all the submodules are combined in a composable way to complete the hardware system.

In short, enabling the CI loop for UVM-based verification and FPGA-based validation has increased the productivity and efficiency of the hardware design flow, through making the verification flow transparent from RTL development. The CI has been also complemented with visual dashboards that collect the relevant information and present the results in a more user-friendly format. The piplines play a key role in the hardware design flow in BZL project for instance, considering the last reporting period (3 weeks) a total 4,536 pipelines were completed, including failed and success ones; in which the duration range for successful pipelines is 2.2 days in average for the largest ones, and 6.4 minutes for the fastest ones.

% --------------------------------------
%\begin{figure}
%  \centering
%  \includegraphics[width=\linewidth]{Figures/CI.jpg}
%  \Description{Illustration showing a test running in parallel on eight FPGAs of a single node. Each FPGA result is marked, where '0' denotes a passed test. The CI system automates deployment scripts.}
%  \caption{A sample test executed in parallel on eight FPGAs of a single node in our FPGA cluster. The job is managed through CI by automating the FPGA deploy scripts as explained in Section~\ref{subsection:fpga2}. ('0' denotes a test passed on that particular FPGA.)}
 % \label{fig:ci}
%\end{figure}
% ---------------------------------

%\input{Conclusion}

\section*{Acknowledgments}
This work has been co-financed by the Barcelona Zettascale Laboratory under project reference REGAGE22e00058408992, with support from the Spanish Ministry for Digital Transformation and Public Services, within the framework of the Recovery and Resilience Facility and the European Union -- NextGenerationEU.

\bibliographystyle{unsrt}
\bibliography{rh_refs}

\end{document}